\documentclass[aps,pre,reprint,showpacs,showkeys,noeprint,superscriptaddress]{revtex4-1} 
\usepackage[utf8]{inputenc}
\usepackage[T1]{fontenc}
\usepackage{amsmath,amssymb,mathtools}
\usepackage{txfonts}

\usepackage{mathrsfs}

\usepackage{graphicx,grffile}
\graphicspath{{/}}

\usepackage{textcomp}
\usepackage{booktabs,tabularx,dcolumn}
\newcolumntype{d}[1]{D{.}{.}{#1}}

\usepackage{url,hyperref}
\usepackage[usenames,dvipsnames]{xcolor}
\hypersetup{colorlinks=true, linkcolor=BrickRed, urlcolor=blue!50!black, citecolor=blue!50!black}

\usepackage[capitalize]{cleveref}
\crefrangelabelformat{equation}{(#3#1#4)$-$(#5#2#6)}

\renewcommand\rho\varrho
\renewcommand\vec[1]{\textrm{\bfseries #1}}

\usepackage[normalem]{ulem}

\begin{document}
\title{When Cooling Meets Flow: Advection-Driven Orientation Selection in Phase Separation around a Moving Cooling Source}

\author{Lakshmipriya K, and Sutapa Roy}
\email{sutapa.roy@hyderabad.bits-pilani.ac.in}
\affiliation{Department of Physics, Birla Institute of Technology and Science, Pilani, Hyderabad Campus, Jawahar Nagar, Kapra Mandal, Medchal District, Telangana 500078, India}

\date{\today}

\begin{abstract}
We investigate phase separation in a binary fluid subjected to a moving Gaussian cooling source and externally imposed material flow. 
Unlike conventional homogeneous quenches, the moving thermal field initiates phase separation at different times across the system, creating a spatially varying thermal history and local “age.” 
The interplay of a moving cooling source, phase separation and external flow field introduces directional bias, resulting in anisotropic pattern formation.
Using an advective Cahn–Hilliard–Cook model coupled to a spatially and temporally varying temperature field, we examine how front propagation, thermal diffusion, phase separation, and material transport jointly select the evolving morphology. 
In the absence of flow, increasing the cooling-source velocity drives a transition from nearly symmetric structures to asymmetric spiral-like patterns and, at higher velocities, to narrow elongated domains as the thermal front increasingly lags behind the source. 
Advection reorganizes the phase-separated domains by transporting, stretching, and reorienting them. 
At low flow velocities, competing thermal and advective histories produce distinct regions of flow-aligned and vertical layers. 
A stronger flow aligns the domains predominantly along the flow direction. 
Our results demonstrate that a moving thermal history can encode a spatial record of non-equilibrium phase separation. 
This work provides a route to controlled morphology selection through the competition between front propagation, thermal diffusion, and material flow.
\end{abstract}

\maketitle

\section{Introduction}

Phase separation is a common phenomena across diverse systems, from alloys, polymers, fluids to biological matter. 
When an initially homogeneous system kept at a temperature higher than the critical demixing temperature $T_c$ \cite{demixing,crctemp} is suddenly quenched below $T_c$, it falls out of equilibrium and moves towards the new equilibrium \cite{func}. 
The intermediate dynamics involves formation and growth of domains of like species. 
This non-equilibrium dynamics is known as phase separation dynamics \cite{bray2002theory,puri2009kinetics,onuki2002phase}.  
Phase separation is well studied in bulk via experiments \cite{oprisan2017,YU2025,merzougui2025}, theory \cite{speck2015,kgwang2023, puri2009kinetics, bray2002theory} and computer simulations \cite{srikanth2026, nalina2024, singh2014, gerhard2024, royepl, puri-nigel, sunil2018}. 
Presence of external fields, confinement, surfaces, anisotropy are known to influence phase separation \cite{wang2000, seul1995, puri1992dynamical}. 
For instance, an imposed external force can cause enhanced collision and coalescence and significantly modify the phase separation kinetics \cite{Chueh2024}. 
Gravity can break the isotropy of domain growth, causing different coarsening behavior \cite{davis2025gravity}. 
Spatial temperature gradients can control the distribution and characteristic sizes of phase-separated domain
\cite{LEE2004,jiang2013}. 
Morphological dependence on shear has also been studied \cite{liu2013}. 
Two-time auto-correlation function has been shown to shed light on the memory effects and dynamics of a phase-segregating fluid \cite{sutapa2019-aging, marco2015}.
However, most of the studies on domain coarsening are based on isothermal systems.

Understanding phase separation under non-isothermal conditions is important because temperature gradients are commonly encountered in the real world. 
Spatially localized thermal fields have been utilized to control material organization and induced phase separation through laser irradiation \cite{gonella2025, ktafi2024,zhou2016laser}. 
Localized optical heating provides another means of controlling phase behavior by spatially modulating temperature, enabling the regulation of biomolecular condensates \cite{condensate2025}. 
Thermally induced phase separation has been employed to fabricate porous polymer composites with tailored morphologies for efficient oil–water separation and adsorption \cite{XU2024}.
By controlling the magnitude and direction of temperature gradients, anisotropic morphologies can be induced and tailored during polymer phase separation \cite{voit2005,Hong_2010}. 
An imposed thermal gradient can produce significant concentration redistribution through thermophoretic transport \cite{duhr2006,amaya2025, alois2019}.

A comprehensive understanding of domain growth in the presence of temperature gradient requires accounting for the evolution of the temporally and spatially varying coupled temperature and composition fields. 
This makes analytical solutions often intractable, thus calling for numerical studies and simulations. 
Monte Carlo simulations were carried out on a two-dimensional Ising ferromagnet in contact with a heat bath that intrinsically has a thermal gradient to explore the critical dynamic behavior \cite{mc-muglia-2012}. 
Molecular dynamics (MD) simulations under temperature gradients are challenging because the imposed heat flux needs to be imposed without violating the hydrodynamic properties of the system. 
The heat exchange algorithm provides an effective approach for maintaining the desired temperature gradient while preserving energy conservation \cite{frenkel2015}. 
Non-equilibrium MD simulations have been employed to study thermodiffusion in binary Lennard-Jones mixtures via calculating Soret coefficient \cite{zimmermann2022}. 
Gonnella et al. studied phase separation in the presence of cold walls and examined the effects of viscosity, composition, and diffusivity on the resulting dynamics using Lattice-Boltzmann (LB) simulations \cite{gonnella-pre2010}. 

Structure formation around moving heating/cooling sources are very interesting and important for technological applications. 
In many manufacturing processes, a localized thermal or processing source moves across a material, so different regions experience different processing conditions over time \cite{liu2024scalable,wang2023,deshmukh2024,furmoto2022,LU2025}. 
Examples include laser processing, additive manufacturing, welding, printing, and coating, where the duration of heating, cooling, or structural evolution can vary from one region to another. 
For example, when a laser spot moves across a powder bed, the local temperature field changes and the material can undergo phase transformation from liquid to solid. 
Simultaneously, the melt flow can transport the material and lead to solidification where the structures depend on the competition between thermal gradients and flow. 
The final microstructure is thus determined not only by the local cooling conditions but by the interplay between thermal-front propagation and material transport. 
A similar history-dependent evolution occurs in inkjet printing, where droplets deposited by a moving print head undergo spreading, internal flow, evaporation, and drying \cite{lohsw2022,chen2023,carou2024,zhang2024}. 
Consequently, material deposited at different times experiences different durations of transport and drying, which ultimately influences the final morphology \cite{bridonneau2020,inkjet2022}. 

Despite its tremendous importance, studies on phase separation under non-isothermal conditions, particularly in the presence of spatially varying temperature fields, are very few and recent. 
Directional quenching by a moving quench boundary can produce distinct domain morphologies, e.g., regular structures oriented parallel or perpendicular to the boundary \cite{kurita2022, kurita2017, furukawa-physica1992}. 
Moving phase-separation fronts can select and control domain orientation and characteristic length scales, providing a way to engineer phase-separated patterns \cite{LBpre2012,foard-pre2009,prl2006biro}. 
Radial propagation of cold temperature fronts from a cooling source moving through a system produces distinct domain morphologies which are controlled by the ratio of the velocities of the cooling source and thermal front \cite{klp2026}.

\begin{figure}
\centering
\includegraphics*[width=0.5\textwidth]{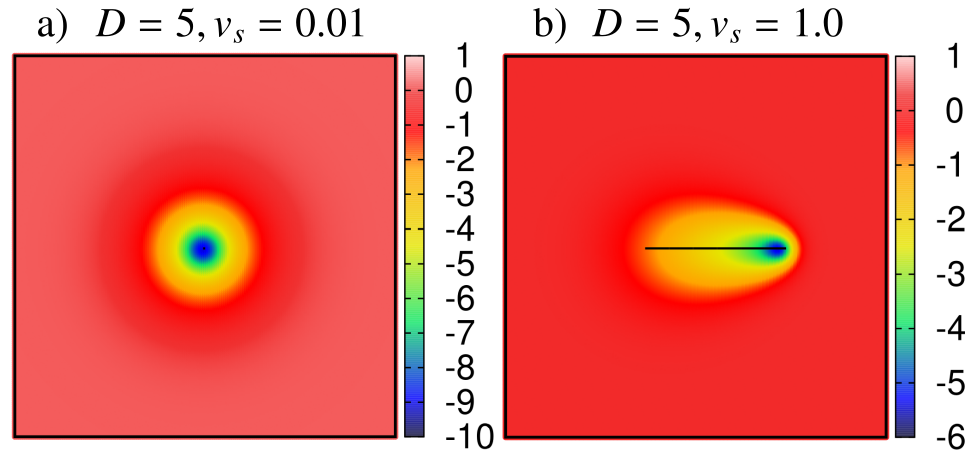}
\caption{Reduced temperature profiles of the system around a moving Gaussian cooling source at fixed $D=5$ for two source velocities: (a) $v_s=0.01$ and (b) $v_s=1$. 
The temperature field $T(r,t)$ is obtained by solving the heat diffusion equation (\cref{heatdiff}).
System parameters are $L=1000$, $T_i=1.2T_c$, $Q=1$ and $\sigma=5$.
}
\label{fig1}
\end{figure}

Presence of advective flow makes the non-equilibrium dynamics more complex and intriguing. 
Advective flow itself does not create domains, but it can change the morphology of existing domains by stretching them, bringing domains together, or separating them. 
Imposed flow is known to drive anisotropic structural reorganization in complex fluids through the interplay of hydrodynamic interactions and interparticle forces \cite{vermant2005}. 
Molecular dynamics (MD) simulations have been used to investigate shear induced phase separation and rheological behavior in three dimensional binary fluids \cite{davis2025}. 
Light-scattering experiments demonstrated that shear flow modifies the growth kinetics of phase separated domains in a direction dependent manner \cite{chan1991}. 
LB simulations reported that oscillatory shear controls phase separation morphology, with nearly isotropic growth at high frequencies and lamellar structures at low frequencies \cite{xu2003}. 
Effects of fluid advection and stirring on phase separation has been studied \cite{onaraigh2008, feng2020}.
A recent study examined enhanced dissipation by advection, highlighting the role of advective transport in accelerating the decay of concentration fluctuations \cite{mazzucato2026}.
Coupling between fluid flow and composition evolution was shown to modify the pathways of domain growth and the resulting morphology \cite{furtado2006}. 
In polymer processing, experiments demonstrated that strong processing flow progressively suppresses phase separation during the downstream evolution of the polymer, while simultaneously influencing the subsequent crystallization process \cite{Ali2022}. 
Nematic interactions and flow or field induced differences in chain tension can be used to drive phase separation in homopolymer melts \cite{zhang2018tension}. 

\begin{figure*}[htbp]
\centering
\makebox[\textwidth]{\includegraphics[width=0.8\paperwidth]{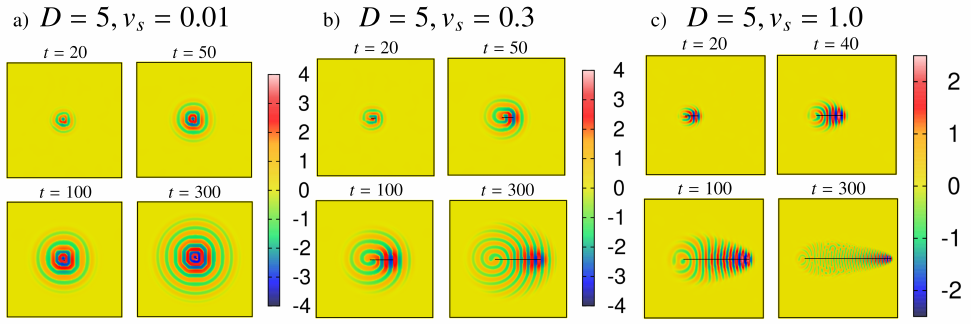}}
\caption{Moving cooling source induced phase separation in a binary mixture for an off-critical composition, $\psi_0=0.1$. 
Time evolution of the temperature and order parameter fields are modeled by the heat diffusion equation and the modified Cahn-Hilliard-Cook equations, respectively. 
Gaussian cooling source is considered. 
Initially the cooling source is centered at $(L/2, L/2)$ and with increasing time it translates along the positive $x$ direction with a constant speed $v_s$. 
The color bar corresponds to the value of order parameter $\psi$, where $\psi>0$ and $\psi<0$ represents A and B rich phases. 
The black solid line represents the source location. 
For clarity, only a portion of the simulation box is displayed.
Results presented are for $L = 1000$, $D=5$, $\sigma=5$, $T_i=1.2T_c$. 
(a) Coarsening patterns for source velocity $v_s=0.01$, at four representative times.
Circular domains with concentric rings are observed in the vicinity of the cooling source. 
Away from the source the system remain in a homogeneous mixed state with $\psi=0$. 
With increasing time, the number of rings increases and the overall domain grows in size. 
(b) Same as (a), but for $v_s=0.3$. 
Initially asymmetric spiral structures are observed and as time progresses semi circular arches form on the right side of the spiral. 
(c) Domain morphology for $v_s=1.0$ at four distinct times. 
At $t=20$ circular layer is observed and as time progresses elongated domain with vertically striped pattern are observed. 
Two alternate layers correspond to opposite phases. 
The source velocity $v_s$ clearly alters the domain morphology. 
}
\label{fig2}
\end{figure*}

Many materials are processed under spatially and temporally varying thermal fields while simultaneously undergoing transport. 
Such coupled thermal and transport effects can significantly influence phase separation dynamics and the resulting morphology \cite{Pooley2005, araki2004}.
Temperature gradients coupled with shear flow can alter phase-separation dynamics, producing distinct morphologies ranging from stripes to droplets \cite{bertei2021}.
Temperature gradients can also induce convection and drive large-scale droplet migration during phase separation \cite{lamorgese2011}.
Thus, understanding how thermal field, phase separation and flow select the domain morphology is important for controlling and engineering material micro-structures to achieve desired properties for specific applications.

In this work, we investigate the phase-separation dynamics of a binary fluid subjected to a moving Gaussian cooling source while the material is simultaneously transported by an imposed fluid velocity. 
The interplay between the phase separation and external driving fields can introduce directional bias, leading to anisotropic phase separation and novel patterns that are absent in conventional quench-induced phase separation. 
We examine flow-controlled orientation selection during phase separation. 
The model captures the generic physics of a moving phase-transition front interacting with material transport, a situation encountered in directional solidification, thermally driven manufacturing processes, printing, coating where local phase evolution occurs under a spatially varying processing history. 

This work is also important from the perspective of `aging' dynamics. 
In conventional phase-separation studies, typically the entire system is quenched at the same time, yielding same aging time throughout the system. 
On the other hand, in our work, the moving cooling source causes different spatial regions to start phase separating from different times. 
Our study is thus relevant for any non-equilibrium proto-typical system in which the thermal history and `age' varies across spatial region.

Rest of the paper is organized as follows: The formulation of the problem and the numerical procedure adopted in this study are described in \cref{model}. 
Results are presented in \cref{results}, where we systematically examine the influence of thermal diffusion, velocity of the moving cooling source, and the imposed flow on the domain morphology and their evolution. 
The main findings are then summarized in the conclusion section \cref{conclusions}, with an outlook and perspective on future
works.

\section{Model and Methods}\label{model}
\begin{figure}
\centering
\includegraphics*[width=0.5\textwidth]{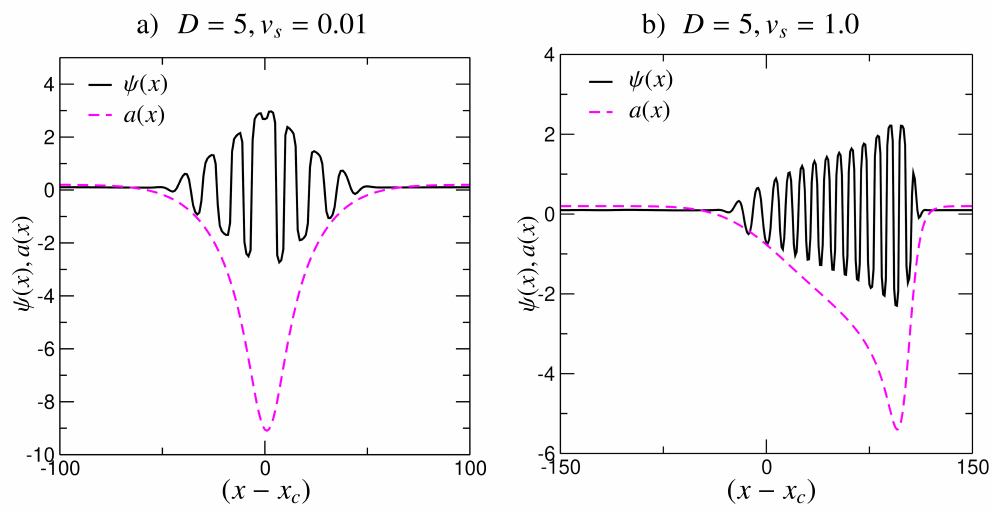}
\caption{Spatial profiles of the order parameter, $\psi(x)$, and the reduced-temperature field, $a(x)$, as functions of the distance from the initial source position, $x_c$, measured along the direction of source motion.
The solid and dashed curves represent $\psi(x)$ and $a(x)$, respectively. 
Parameter values for (a) and (b) are same as \Cref{fig2}(a) and (c), respectively. 
Domains are confined to the spatial region with $a(r,t)<0$. 
Whereas the region with $a(x)>0$ remains in the homogeneous mixed state, with $\psi(x)\simeq\psi_0$.
}
\label{fig3}
\end{figure}

We investigate non-equilibrium phase separation dynamics in a two-dimensional binary fluid subjected to a moving cooling source and an imposed advective flow. 
Initially, the fluid is maintained above its critical demixing temperature $T_c$ \cite{crctemp, MEFisher1967, demixing}, while the source is held below $T_c$. 
As the source moves, it creates a time-dependent temperature field $T(\vec r,t)$ that drives the evolution of the order-parameter field $\psi(\vec r,t)$. 
Phase separation is favored in regions where $T<T_c$. 
For Binary fluids the order parameter $\psi$ is the concentration difference between A and B species \cite{bray2002theory}. 

Time evolution of $T(\vec r,t)$ is modeled by the heat diffusion equation \cite{sroy2018,Rc1990} given by
\begin{equation}
\label{heatdiff}
\frac{\partial T}{\partial t}
= D\nabla^2 T
-Q\exp\left[-\frac{(x-x_c(t))^2+(y-y_c(t))^2}{2\sigma^2}\right]
\end{equation}
The Gaussian function here describes the cooling source which is centered at $(x_c(t),y_c(t))$ at time $t$. 
Here, $Q$ and $\sigma$ denote the amplitude and spatial width of the Gaussian. 
The source propagates through the system with a velocity $v_s$. $D$ is the diffusivity of the material. 

To describe the evolution of the composition field and associated phase separation dynamics, we employ the modified Cahn–Hilliard–Cook (CHC) framework \cite{benjamin1999, cahn1958, cook1970}  with an advective contribution. 
The formulation is derived from the free energy of the system, expressed in terms of the Landau–Ginzburg free energy functional \cite{func, bray2002theory},
\begin{equation}
\label{free}
\frac{F}{k_BT_c} = \frac{1}{\nu_0} \int d^{3}r \Bigl[ \frac{1}{2} a{\psi}(r)^{2}
+ \frac{1}{4} b{\psi}(r)^{4} + \frac{1}{2} C\bigl({\nabla} {\psi}(r)\bigr)^{2}\Bigr],
\end{equation} 
where, $F$ is the free energy, $\nu_0$ is a volume unit and $a<0$, $b>0$, $C>0$ are constants. 
$a \propto (T-T_c)/T_c$ dictates deviation from the critical temperature $T_c$ \cite{Roy_2011}. 
For $a<0$ one would thus get phase separation. 
We consider order parameter conservation through the continuity equation,  
\begin{equation} \label {chc1}
 \frac{\partial{\psi}(r,{t})}{\partial {t}}+{\nabla} \cdot {\boldsymbol j}({r},{t})=0, 
\end{equation}
where, $j$ is the current density ${\boldsymbol j}({r},{t})= -M \nabla \mu $. 
$M$ is the mobility and $\mu=\partial F / \partial \psi$ is the chemical potential.
Substituting the values, we obtain 
\begin{equation}
    \label{chc} 
   \begin{aligned}
\frac{\partial \psi(\mathbf{r},t)}{\partial t}
&= -\nabla\cdot\mathbf{j}(\mathbf{r},t)
 = \nabla\cdot\left(M\nabla\mu\right) \\
&= M\nabla^2\mu = M\nabla^2\left(\frac{\delta F}{\delta\psi}\right) 
\end{aligned}
\end{equation}
To attain the dimensionless form of the governing equation, we introduce the following rescaling factors:
\begin{equation}
\label{rscal}
     \tilde{\psi} = \sqrt{\frac{a}{b}},\quad
       \tilde{r} = \sqrt{\frac{C}{a}},\quad
    \tilde{t} = \frac{\nu_0\tilde{r}^2}{Mk_BT_c a}
\end{equation}
Using these rescaling factors in \cref{free,chc}, the dimensionless Cahn-Hilliard-Cook (CHC) equation takes up the form
\begin{equation}
\label{scaledchc}
\frac{\partial \psi(\vec r,t)}{\partial t} = \nabla ^2 \Big ( a \psi(\vec r,t) + \psi^3(\vec r,t) - \nabla^2 \psi(\vec r,t) \Big) 
\end{equation}
In the above formalism (\cref{scaledchc}), $a$ is a negative constant. 
In our system, due to heat diffusion $a$ is no longer a constant, rather it becomes a time and space dependent field $a(\vec r,t)$. 
Thus, we replace $a$ in (\cref{scaledchc}) by $a(\vec r,t)$. 
Note that $a(\vec r,t)=a_0(T(\vec r,t) -T_c)/T_c$, $a_0$ being a positive constant. 
For a particular model system, $T_c$ is constant. 
If $T(r,t)>T_c$, $a(r,t)>0$, corresponding to the mixed state. 
While, for $T(r,t)<T_c$, $a(r,t)<0$, drives the system into the phase separated state.  

The modified CHC equation thus becomes
\begin{equation}
\label{chc2}
\frac{\partial \psi(\vec r,t)}{\partial t} = \nabla ^2 \Big ( a(\vec r,t) \psi(\vec r,t) + \psi^3(\vec r,t) - \nabla^2 \psi(\vec r,t) \Big) + \eta(\vec r,t)
\end{equation}
Here, the first term on the RHS takes care of the direct coupling between the temperature and the composition fields. 
We have added a Gaussian random noise $\eta(\vec r,t)$ which follows the relation $\langle \eta(\vec r,t)\eta(\vec r',t') \rangle= -2\nu \nabla^2\delta(\vec r - \vec r')\delta(t - t') $, where $\nu(\vec r)$ is the noise strength \cite{roy2018colloid, sroy2018}. 
The time-dependent temperature field $(T(\vec r,t)$ is obtained by solving \cref{heatdiff} and is fed into \cref{chc2} through $a(\vec r,t)$.

In the presence of advection, the phase separating domains are simultaneously transported by the fluid flow. 
Consequently, the evolution of the composition field is governed by the competition between phase separation induced by the moving thermal source and advection due to the fluid velocity $\mathbf u$. 
The imposed flow enters the modified order-parameter equation by adding a coupling term $(\mathbf u\cdot \nabla \psi)$ \cite{liu2013,scarpa2021,feng2020} to \cref{chc2}.

\begin{figure}
\centering
\includegraphics*[width=0.5\textwidth]{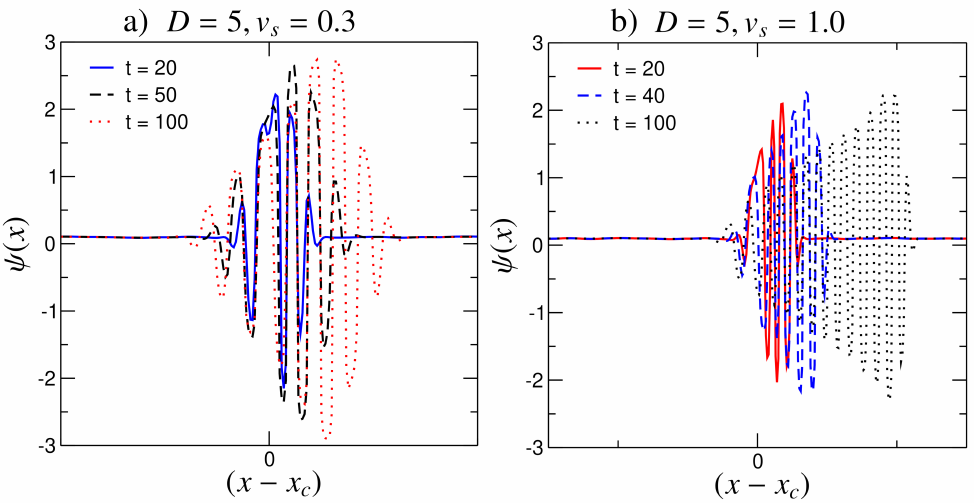}
\caption{Plot of the local order parameter profile $\psi(x)$ along the direction of source motion at three distinct times. 
(a) and (b) correspond to $v_s=0.3$ and $v_s=1$, respectively. 
For both cases, $D=5$, $L=1000$, $\sigma=5$, $T_i=1.2T_c$, and the source is initially placed at $(L/2,L/2)$.
}
\label{fig4}
\end{figure}

For our numerical calculations, initially ($t<0$) the system is maintained at a high temperature $T_i=1.2 T_c$. 
At $t=0$, the Gaussian cooling source is placed within the simulation box at the location $(L/2,L/2)$, $L$ being the length of the square box. 
The source moves along the positive $x$- axis with a velocity $v_s$. 
Width and amplitude of the Gaussian function are kept constant at $\sigma=5$ and $Q=1$. 
Periodic boundary conditions \cite{allen2002computer} are applied along all Cartesian directions. 
The relevant equations are numerically solved using the finite difference method \cite{butcher2008numerical} with a time step $dt=0.001$. 
All results presented here are for a simulation box size $L=1000$ and an average order parameter $\psi_0=0.1$. 

\begin{figure*}[htbp]
\centering
\makebox[\textwidth]{\includegraphics[width=0.8\paperwidth]{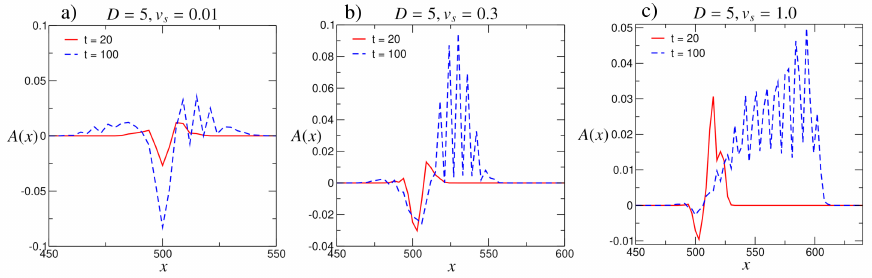}}
\caption{Anisotropy parameter $A(x)$, calculated from \cref{Aeqn}, for three different values of $v_s$ and fixed $D$. 
All other system parameters are same as \cref{fig1}. 
For domains aligned along the vertical and horizontal directions, $A>0$ and $A<0$, respectively. 
Conversely, $A\simeq 0$ for circular domains. 
(a), (b), (c) correspond to $v_s=0.01$, $0.3$, and $1$, respectively. 
Nonzero values of $A$ occur only within the region where phase-separated domains have developed. 
For $v_s=0.01$, the spatially averaged value of $A$ remains close to zero, confirming the presence of circular domains. 
For (b) and (c), the pronounced positive peaks in $A(x)$ indicate the prevalence of vertically oriented layers. 
Thus, the spatial and temporal evolution of $A(x)$ provides a useful quantitative measure of anisotropy in the phase-separated patterns.
}
\label{fig5}
\end{figure*}

\begin{figure}
\centering
\includegraphics[width=0.3\textwidth]{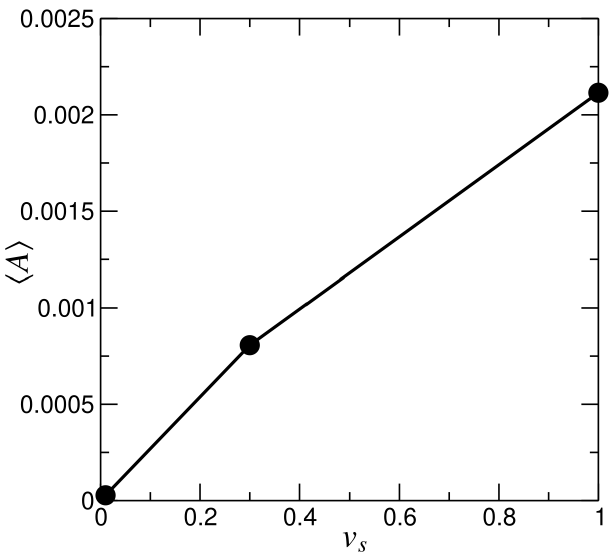}
\caption{Plot of spatially-averaged anisotropy parameter, $\langle A\rangle$, as a function of the source velocity $v_s$. 
System parameters are $L=1000$, $D=5$, $\sigma=5$, and $T_i=1.2T_c$. 
The monotonic increase in $\langle A\rangle$ with $v_s$ demonstrates that faster source motion enhances the anisotropy and promotes the preferential vertical alignment of the phase-separated domains.
}
\label{fig6}
\end{figure}

\section{Results}\label{results}

\subsection{Without Flow:}\label{without flow}

Understanding pattern morphology and their evolution is crucial to understanding the dynamics of phase separation. 
We first investigate the temperature field generated by the moving Gaussian cooling source. 
Results presented in \cref{fig1} correspond to an off-critical mean composition of $\psi_0=0.1$, with the source moving along the positive $x$-direction, for two distinct values of source velocity $v_s$.
It represents the reduced temperature profiles obtained for $D=5$ at two different source velocities, $v_s=0.01$ and $1$. 
The system size and source parameters are fixed at $L=1000$, $T_i=1.2T_c$, $\sigma=5$, and $Q=1$.
The temperature field is evolved using the heat diffusion equation \cref{heatdiff}. 
For a lower source velocity, $v_s=0.01$, the cooling source moves only a small distance over the time. 
Consequently, the temperature profile retains an approximately circular symmetry.
As the source velocity is increased to $v_s=1$ (\Cref{fig1}(b)), the source travels farther along the positive $x$-direction. 
The increased source motion breaks the circular symmetry of the temperature field, resulting in a noticeable asymmetric profile.

\begin{figure}
\centering
\includegraphics*[width=0.4\textwidth]{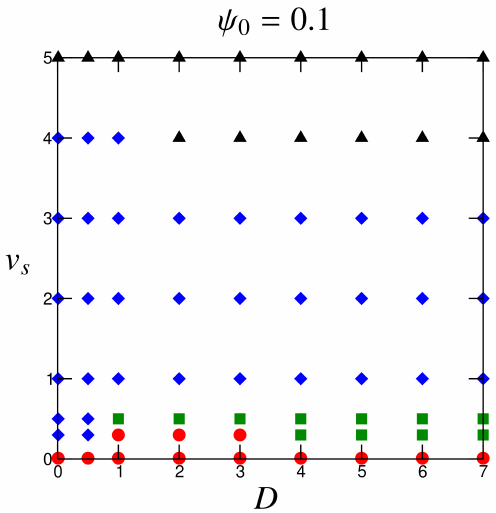}
\caption{Phase diagram of the system in the $D-v_s$ parameter plane. 
All data refer to system size $L=1000$ and initial temperature $Ti = 1$. 
Symbols correspond to distinct pattern shapes: Circle, concentric circular rings; square, spiral with semicircular arches pattern; diamond, elongated domain with vertical stripes; triangle, elongated domain with highly asymmetric structure.}
\label{fig7}
\end{figure}

\begin{figure*}
\centering
\includegraphics[width=0.7\paperwidth]{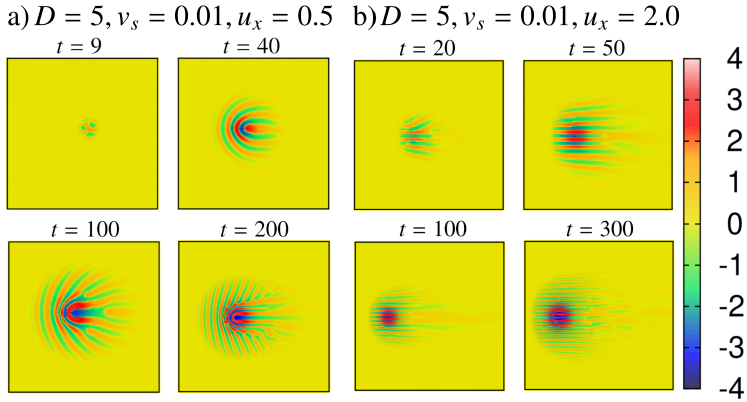}
\caption{Phase separation snapshots of a binary mixture with an off-critical composition, $\psi_0=0.1$, in the presence of fluid flow. 
Initially, the Gaussian cooling source is centered at $(L/2,L/2)$ and moves in the positive $x$ direction with a constant velocity $v_s$, while the fluid flows with velocity $\mathbf u=(u_x,0)$.
(a) Phase separation patterns for parameters $D=5,v_s=0.01,u_x=0.5$ at four representative times. 
Regions that enter the phase-separation regime first are transported by the imposed flow and gradually reorient into predominantly horizontal domains. 
In contrast, regions entering phase separation later experience the flow for a shorter duration and evolve into vertically oriented stripes. 
(b) Coarsening patterns for $D=5,v_s=0.01,u_x=2.0$ at four representative times. 
With increasing fluid velocity, advection dominates, orienting the phase-separated domains along the flow direction and producing horizontal stripes. 
}
\label{fig8}
\end{figure*}

Next, we examine the pattern morphology and its dependence on $v_s$ in the absence of advection. 
Figure \ref{fig2} shows the coarsening snapshots for different values of $v_s$ and fixed $D=5$. 
The color bar represents the full range of order parameter $\psi$, where, $\psi>0$ and $\psi<0$ denote phases A and B, respectively. 
Figure \ref{fig2}(a) corresponds to $v_s=0.01$ and the phase separation patterns are shown at four representative times. 
Circular domains with concentric rings are observed. 
Note that at this small velocity the source behaves approximately like a stationary source and the resulting morphology retains circular symmetry.
At $t=20$, a ring of phase B is formed near to the initial source position, followed by a ring of phase A. 
As time progresses, more rings of alternating phases develop and the overall domain size grows.
Away from the cooling source, the system remains in the homogeneous phase with $\psi=0$. 
For better clarity, only a portion of the simulation box is displayed. 

Next, in \Cref{fig2}(b), the source velocity is increased to $v_s=0.3$. 
In contrast to the circularly symmetric morphology observed in \Cref{fig2}(a), the resulting patterns exhibit pronounced asymmetry and develop into spiral-like structures.
At later times ($t=100$ and $t=300$), semicircular arches form ahead of the spiral as the source moves toward the right.
Successive layers correspond to alternating A-rich and B-rich phases.
The solid black line indicates the trajectory of the moving source. 
Thus, increasing $v_s$ breaks the circular symmetry of the domains and promotes their preferential alignment along the direction of source motion.

The source velocity is further increased to $v_s=1$, and the corresponding domain morphologies at four representative times are shown in \Cref{fig2}(c).
At early time ($t=20$), the pattern consists of circular layer near the source.
As the system evolves, elongated domains with vertical stripes form. 
With increasing time, the number of stripes and the domain size increase. 
For larger $v_s$ we observed highly asymmetric structures (not presented here). 

To elucidate the mechanism of pattern formation, \Cref{fig3} shows the spatial profiles of the order parameter, $\psi(x)$, and the reduced-temperature field, $a(x)$, as functions of the distance from the initial source position, $x_c$, measured along the direction of source motion.
The solid and dashed curves represent $\psi(x)$ and $a(x)$, respectively. 
The oscillations in $\psi(x)$ correspond to the alternating A-rich and B-rich layers observed in \Cref{fig2}.
Parameter values for \Cref{fig3}(a) and (b) are same as \Cref{fig2}(a) and (c), respectively. 
Consequently, the symmetric and asymmetric morphologies observed in these two cases are also reflected in the corresponding order-parameter profiles.

A negative value of $a(x)$ corresponds to a local temperature below the critical temperature, $T(x)<T_c$, and therefore favors phase separation.
As observed in \cref{fig3}, the domains are confined to the spatial region with $a(r,t)<0$. 
Whereas the region with $a(x)>0$ remains in the homogeneous mixed state, with $\psi(x)\simeq\psi_0$.
On the other hand, Soret effect is known to lead to domain formation in regions with $T>T_c$ as well \cite{Onsager1931,KöhlerMorozov2016,soret}. 
In our case, clearly phase separation is governed predominantly by the motion of the cooling source, not Soret effect and is `local' in nature. 

\begin{figure*}
\centering
\includegraphics[width=0.8\paperwidth]{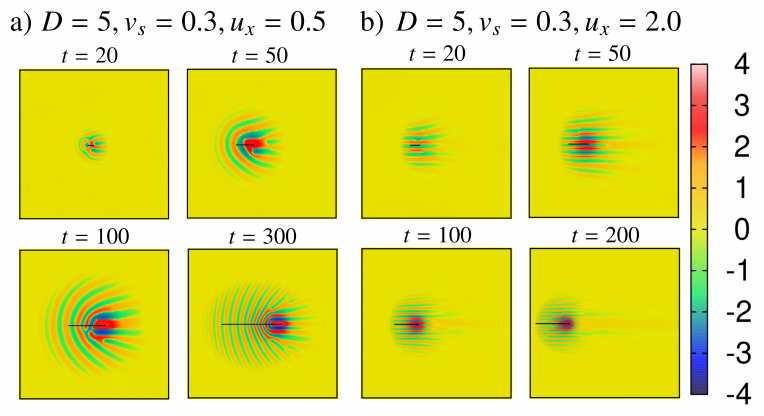}
\caption{(a) Demixing morphology for $D=5,v_s=0.3,u_x=0.5$ at four distinct times. 
The fluid flow disrupts the spiral structure, reorienting it horizontally, while the immediately quenched region becomes vertical. 
(b) Phase separated snapshots for $D=5,v_s=0.3,u_x=2.0$ at four characteristic times.
The patterns orient along the flow direction, parallel to the $x$-axis, throughout the evolution.
}
\label{fig9}
\end{figure*}

To quantify the domain morphology, we examine the local order parameter profile $\psi(x)$ along the direction of source motion. 
\Cref{fig4}(a) presents $\psi(x)$ at three representative times for $D=5$ and $v_s=0.3$; the corresponding morphological snapshots are presented in \Cref{fig2}(b).
The peaks and troughs in $\psi(x)$ arise from the layers of the two phases (see snapshots). 
At early time $t=20$, $\psi(x)$ is almost symmetric around $x=x_c$. 
At later times ($t=50$ and $100$), additional layers develop predominantly to the right of the initial source position ($x>x_c$), resulting in an increasingly asymmetric profile.
\Cref{fig4}(b) presents the corresponding profiles for $D=5$ and $v_s=1$.
Owing to the higher source velocity, the profile is more strongly asymmetric and extends farther in the positive $x$-direction.
For both cases, system parameters are $L=1000$, $\sigma=5$, $T_i=1.2T_c$, and the source is initially placed at $(L/2,L/2)$.

Note that this system is particularly interesting because different parts of the system start phase separating from different times. 
Under such conditions, the assumption of a single self-similar growing length scale underlying conventional dynamic scaling might not work \cite{sroy2013,sroy2012, bray2002theory}.

To elucidate the origin of the morphological asymmetry and the preferential orientation of the phase-separated layers, we quantify the degree of anisotropy using the anisotropy parameter $A$, defined as
 \begin{equation} \label{Aeqn}
      A=\left \langle \left (\frac{\partial \psi}{\partial x} \right)^{2} \right \rangle - \left \langle \left (\frac{\partial \psi}{\partial y} \right )^{2} \right \rangle
 \end{equation}
For domains aligned along the vertical direction, variations in $\psi$ occur predominantly along the $x$-direction, such that $\lvert\partial\psi/\partial x\rvert$ is relatively large and $A>0$. 
Conversely, horizontally aligned domains exhibit stronger variations along the $y$-direction, resulting in $A<0$. 
For circular symmetric domains, $A$ is expected to be zero. 
For computational purposes, the simulation domain is partitioned into multiple vertical bins of same width, and $A$ is evaluated separately within each bin to obtain the spatially resolved anisotropy profile $A(x)$.

\Cref{fig5}(a) shows the spatial profile $A(x)$ at two representative times for $D=5$ and $v_s=0.01$. Nonzero values of $A$ occur only within the region where phase-separated domains have developed. 
Each peak in $A(x)$ corresponds to an individual layer in the order-parameter profile $\psi(x)$.
At the early time $t=20$, $A(x)$ exhibits positive and negative peaks of relatively small amplitude, indicating that the contributions from order-parameter gradients along the $x$- and $y$-directions are nearly balanced. 
This behavior is consistent with the approximately circular symmetry of the morphology observed for $D=5$ and $v_s=0.01$ in \Cref{fig2}(a). 
At the later time $t=100$, the magnitudes of the local peaks increase; nevertheless, the spatially averaged value of $A$ remains close to zero, confirming the absence of a preferred domain orientation.

\Cref{fig5}(b) shows the spatial variation of $A(x)$ for $D=5$ and $v_s=0.3$. The low-amplitude negative values of $A(x)$ originate from the horizontally oriented segments of the spiral, whereas the pronounced positive peaks arise from the vertical layers that develop to its right [see \Cref{fig2}(b)]. At the higher source velocity $v_s=1$ [\Cref{fig5}(c)], positive values of $A(x)$ become increasingly dominant because of the prevalence of vertically oriented layers. Thus, the spatial and temporal evolution of $A(x)$ provides a useful quantitative measure of anisotropy in the phase-separated patterns.

\Cref{fig6} shows the system-averaged anisotropy parameter, $\langle A\rangle$, as a function of the source velocity $v_s$. The monotonic increase in $\langle A\rangle$ with $v_s$ demonstrates that faster source motion enhances the anisotropy and promotes the preferential vertical alignment of the phase-separated domains.

To systematically characterize the dependence of the domain morphology on the governing parameters, in \Cref{fig7} we present the phase diagram of the system in the $D-v_s$ parameter space. 
Symbols denote different domain morphologies observed. 
Circle: concentric circular rings; square: spiral with semicircular arches; diamond: elongated domain with vertical stripes; triangle: elongated domain with highly asymmetric structure.

\subsection{With Flow:}

\begin{figure*}
\centering
\includegraphics[width=0.8\paperwidth]{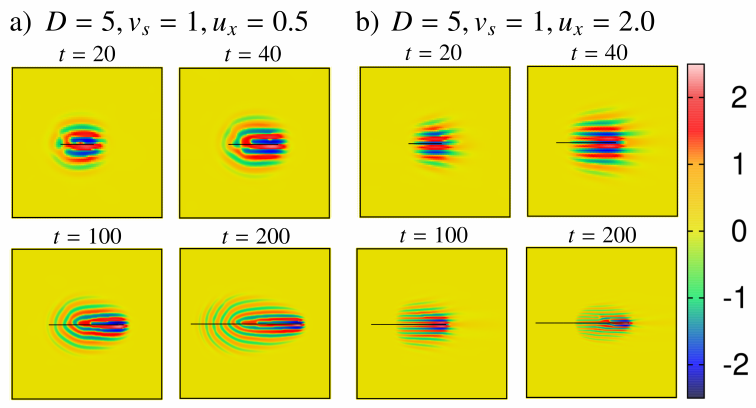}
\caption{(a) Coarsening patterns for $D=5,v_s=1.0,u_x=0.5$.
The immediately quenched region forms a \textit{c-shaped} structure, while the longer-quenched region exhibits horizontal stripes. 
(b) Snapshots of the demixing patterns for $D=5$, $v_s=1.0$, and $u_x=2.0$. 
The black solid line represents the source propagation. 
For clarity only a portion of the box is displayed.
}
\label{fig10}
\end{figure*}

We next analyze the phase separation dynamics when the system is subjected to fluid advection
Note that without the flow term, only the thermal source moves and the cooling spot passes over the material. 
As a result, domains form in spatial regions where the temperature is below the critical temperature. 
In the presence of flow, these domains get transported as well. 
Exploring the ensuing dynamics is thus interesting as it captures how a moving fluid transports the phase-segregated structures that are formed by the moving cooling spot while phase separation is still ongoing. 

The resulting dynamics is investigated by systematically varying the flow velocity along the horizontal direction, $u_x$, while setting the transverse component to zero, $u_y=0$. 
Unless stated otherwise, the simulations are performed using $L=1000$, $dt=0.001$, $\sigma=5$, and $T_i=1.2T_c$.

\Cref{fig8}(a) shows the evolution of the phase-separating morphology at four representative times for $D=5$, $v_s=0.01$, and a horizontal flow velocity $u_x=0.5$. 
The color scale spans the full range of the order parameter. 
For clarity, only a portion of the simulation domain is displayed. 
For comparison, in the absence of flow ($u_x=0$), the same set of parameters produces symmetric concentric rings [see \Cref{fig2}(a)]. 
The imposed flow substantially alters the domain morphology.
At very early time, circular rings form. 
However, this circular shape is progressively disrupted by advection, leading to increasingly complex morphologies.

Because the cooling source moves through the system, different spatial regions start phase separation at different times. 
Once the local temperature falls below the critical temperature, such that $a(\vec{r},t)<0$, domains begin to form and grow within that spatial region. Simultaneously, the emerging domains are advected along the positive $x$-direction and gradually align with the imposed flow.

The resulting morphology can therefore be divided into two characteristic regions. (i) Farther downstream, toward the right, the domains have undergone phase separation and advection for a relatively long time and consequently exhibit a predominantly horizontal orientation. 
(ii) In contrast, regions closer to the left have started phase separation more recently and the domains have experienced advective flow for a shorter duration. These domains retain more of their source-induced morphology and may therefore remain vertically oriented rather than aligning horizontally. 
Thus, the spatially varying morphology reflects the different quench times and flow-exposure durations across the system. 
Far from the cooling source, the system remains in the homogeneous mixed state, with $\psi\simeq\psi_0$.

\Cref{fig8}(b) shows the domain morphology at four representative times for a higher flow velocity, $u_x=2$, while keeping $D=5$ and $v_s=0.01$, as in \Cref{fig8}(a). 
At this higher value of $u_x$, advection plays a more dominant role, producing a stronger alignment of the phase-separated domains with the horizontal flow direction. 
Consequently, vertically oriented stripes are no longer observed. 
As the system evolves, more number of horizontally aligned layers develop. 
The solid black line indicates the trajectory of the moving Gaussian cooling source.

To examine the influence of the source velocity on phase-separation dynamics, \Cref{fig9} presents the morphological evolution for the higher velocity $v_s=0.3$, with $D=5$ and flow velocities $u_x=0.5$ and $2$, as considered in \Cref{fig8}. 
By comparison, in the absence of flow ($u_x=0$), the same source velocity produces spiral patterns [see \Cref{fig2}(b)]. 
The introduction of advection disrupts these spiral structures and progressively reorients the domains along the horizontal flow direction.
For $u_x=0.5$ [\Cref{fig9}(a)], the newly quenched region initially develops C-shaped domains, which subsequently evolve into vertically oriented structures. In contrast, at the higher flow velocity $u_x=2$ [\Cref{fig9}(b)], advection dominates the morphological evolution, and only horizontally aligned layers are observed.
 
\Cref{fig10} presents the morphological evolution for the higher source velocity $v_s=1$. 
The qualitative influence of the flow velocity $u_x$ is similar to that observed for $v_s=0.3$ in \Cref{fig9}, with advection promoting domain alignment along the flow direction. 
Regions far from the cooling source remain in the homogeneous mixed state. 
The solid black line indicates the trajectory of the moving Gaussian cooling source. 
Overall, the phase-separation dynamics is governed by the competition among thermal-front diffusion, motion of the localized cooling source, and advective transport.

\begin{figure}
\centering
\includegraphics*[width=0.5\textwidth]{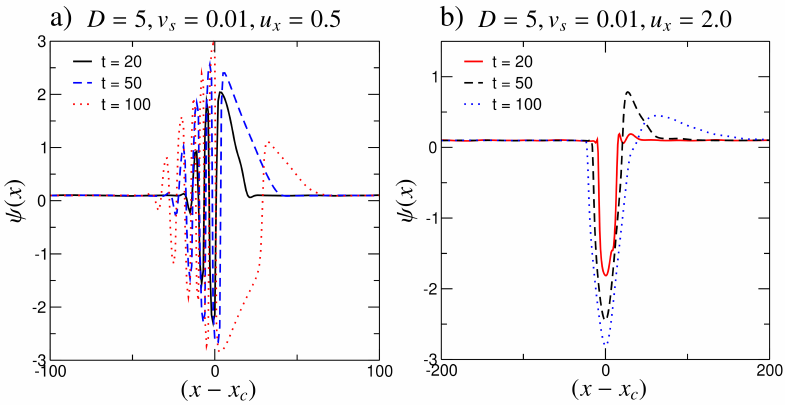}
\caption{Variation of the order parameter $\psi(x)$ with the relative distance $(x-x_c)$ along the source-propagation direction for $D=5$ and $v_s=0.01$, shown at three representative times in the presence of an imposed flow with velocity $u_x$. 
(a), (b) correspond to imposed fluid velocity $u_x=0.5$ and $u_x=2.0$, respectively. 
}
\label{fig11}
\end{figure}

To quantify the domain morphology, we calculate the order parameter profile $\psi(x)$ along the direction of the source motion. 
\Cref{fig11} presents $\psi(x)$ at three representative times for $D=5$, $v_s=0.01$ and two distinct $u_x$. 
For both cases, system parameters are $L=1000$, $\sigma=5$, $T_i=1.2T_c$, and the source is initially placed at $(L/2,L/2)$.
The peaks in $\psi(x)$ arise from the layers of the two phases (see snapshots). 
\Cref{fig11}(a) presents the corresponding profiles for $u_x=0.5$
In comparison with the order parameter profile with $u_x=0$, the symmetry in the profile is lost. 
At late times, patterns extend towards the right and the $\psi(x)$ profile is also non-zero over an extended region for $x>x_c$. 

\begin{figure}
\centering
\includegraphics*[width=0.5\textwidth]{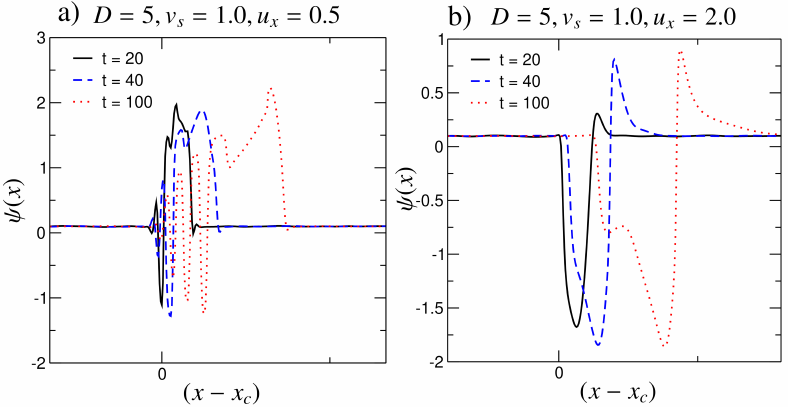}
\caption{(a) Order parameter profile $\psi(x)$ as a function of the relative distance $(x-x_C)$ calculated along the direction of source propagation for $D=5$, $v_s=1.0$, and fluid velocity $u_x=0.5$ at three representative times. 
(b) Same parameters as (a), with $u_x=2.0$.
}
\label{fig12}
\end{figure}

The variation of order parameter along the direction of source propagation for $D=5, v_s=1.0,$ with $u_x=0.5$ and $u_x=2.0$ is presented in \Cref{fig12}(a) and \Cref{fig12}(b) respectively, at three selected times. 
For $u_x=0.5$, the profiles exhibit alternating phase regions, with a progressive rightward shift with increasing time due to the fluid flow from left to right. 
A similar rightward displacement is observed for $u_x=2.0$. 
Fewer layers are present due to the predominantly horizontal alignment of the phase separated structures. 

Advection does not itself induce the formation of domains; rather, it modifies their subsequent evolution by stretching and rotating the domains. 
To elucidate the origin of the morphological asymmetry and the preferential orientation of the phase-separated layers, we quantify the degree of anisotropy using the anisotropy parameter $A(x)$ defined in \cref{Aeqn}. 

\begin{figure}
\centering
\includegraphics*[width=0.5\textwidth]{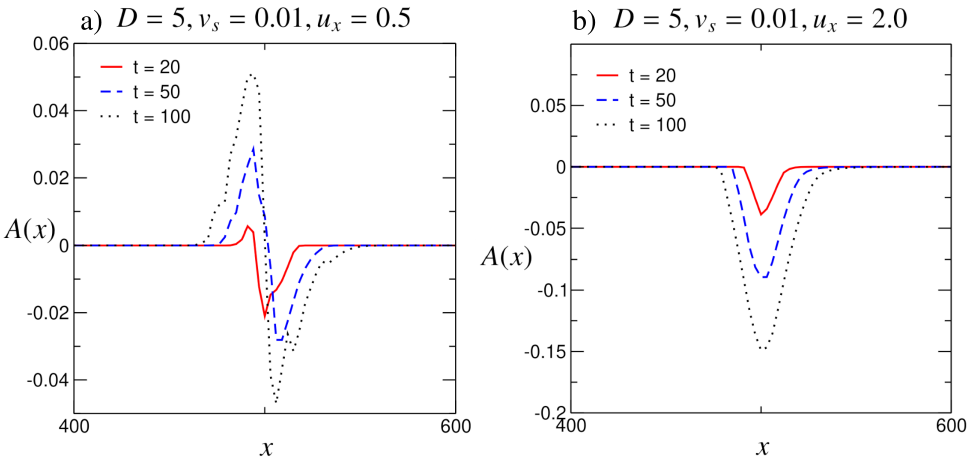}
\caption{(a) Anisotropy parameter $A(x)$ as a function of $x$ for $D=5$, $v_s=0.01$, $u_x=0.5$, at three distinct times.
Positive and negative peaks in $A$ correspond to vertically and horizontally oriented stripes, respectively. 
(b) $A(x)$ profile for  $D=5$, $v_s=0.01$, and $u_x=2.0$ at three representative times. 
The predominantly negative $A$ confirms the horizontal stripe morphology.
}
\label{fig13}
\end{figure}

In \Cref{fig13}(a), the anisotropy parameter $A(x)$ is presented as a function of $x$ for parameters $D=5,v_s=0.01,u_x=0.5$. 
Nonzero values of $A$ occur only within the spatial region where phase-separation occurred.
The profile of $A$ displays both positive and negative peaks, reflecting the coexistence of differently oriented phase separated structures in the system. 
The \Cref{fig13}(b) illustrates the spatial variation of $A(x)$ at three representative times for $u_x=2.0$. 
Other parameters are same as \Cref{fig13}(a).
For this large $u_x$, $A(x)$ becomes negative, reflecting the preferential horizontal alignment of the domains, consistent with the observed stripe morphology.

\begin{figure}
\centering
\includegraphics*[width=0.5\textwidth]{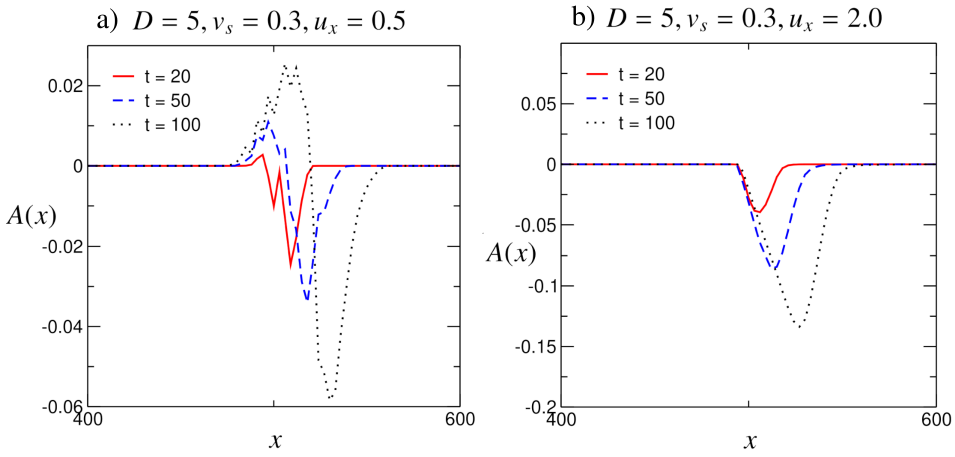}
\caption{The variation of $A(x)$ with $x$ at three representative times for $D=5$, $v_s=0.3$. 
(a) The imposed fluid velocity is $u_x=0.5$, the profile has both positive and negative peaks. 
(b) The same plot for $u_x=2.0$.  
The profile exhibits pronounced negative troughs, indicating a dominance of $y$-directional gradients.
}
\label{fig14}
\end{figure}

The same is plotted for a larger $v_s=0.3$ in \Cref{fig14}(a) and (b) for $u_x=0.5$ and $u_x=2.0$, respectively. 
$A(x)$ follow a trend broadly similar to that observed in \Cref{fig14}.  
However, a rightward shift is observed here because of the increased source velocity.  

\begin{figure}
\centering
\includegraphics*[width=0.5\textwidth]{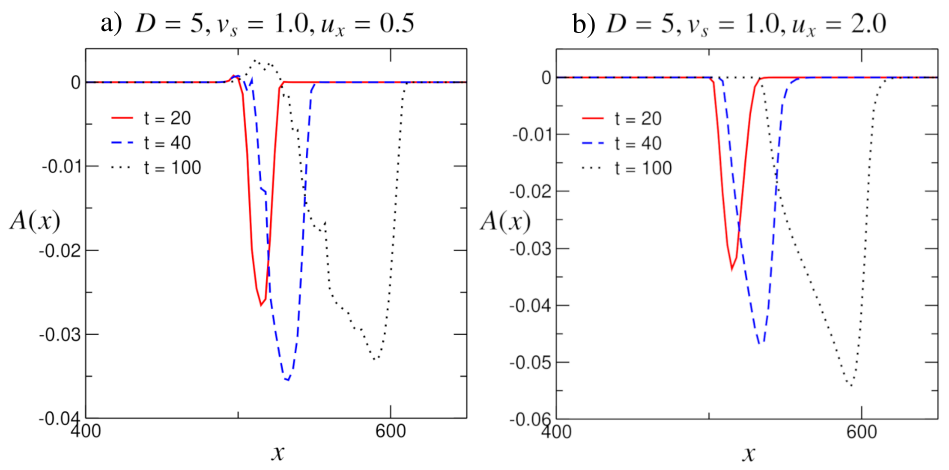}
\caption{(a) Plot of $A(x)$ vs. $x$ for $D=5,v_s=1.0$ and imposed fluid velocity $u_x=0.5$ at three different times. 
(b) $A(x)$ profile for $D=5,v_s=1.0,u_x=2.0$. 
The system parameters are $L=1000, dt=0.001, \sigma=5,$ and initial temperature $T_i=1.2T_c$.
}
\label{fig15}
\end{figure}

\Cref{fig15}(a) demonstrates the $A(x)$ profile for $D=5,v_s=1.0,u_x=0.5$ at three selected times. 
The plot shows prominent negative peaks even for relatively low fluid velocity $v_s=0.5$.
In addition, a comparatively weak positive peak appears, which is associated with the \textit{c-shaped} structure.
For $u_x=2.0$, the anisotropy parameter $A$ remains negative as depicted in \Cref{fig15}(b), which is coherent with the formation of horizontally oriented domain structures. 
Furthermore, as time progresses, the troughs in the $A$ profile shift progressively toward the right, reflecting the increasing influence of the imposed fluid flow and source velocity $v_s$ along the positive $x$ direction.

\begin{figure*}[htbp]
\centering
\makebox[\textwidth]{\includegraphics[width=0.8\paperwidth]{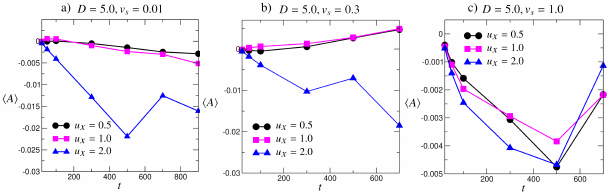}}
\caption{Spatially averaged anisotropy parameter $\langle A\rangle$ as a function of time for fixed $D=5$ and three different fluid velocities $u_x$. 
(a) Source velocity $v_s=0.01$. 
(a) Source velocity $v_s=0.3$. 
(b) Source velocity $v_s=1.0$. 
Different symbols denote the respective values of $u_x$.
}
\label{fig16}
\end{figure*}

Next, in \Cref{fig16}, we present the system-averaged anisotropy parameter $\langle A\rangle$ as a function of time for different values of fluid velocity $u_x$. 
Different symbols correspond to different $u_x$. 
Across (a)-(c), the source velocity $v_s$ increases. 
The qualitative trend is similar for all of them. 
At early times $\langle A \rangle$ can exhibit positive peaks. 
However, at late times as the system evolves $\langle A \rangle$ progressively shifts toward large negative values, reflecting the development of horizontal layers parallel to the $x$ axis.  
Magnitudes close to zero indicate the presence of nearly isotropic morphologies. 
The imposed flow clearly controls the pattern morphology. 

\section{Conclusions}\label{conclusions}

In conclusion, we study how a moving thermal front and material transport jointly select morphology during phase separation in a binary fluid. 
Due to heat diffusion the system contains spatially and temporally varying temperature and composition fields. 
Coupling between these two fields dictate the coarsening dynamics. 
Without fluid flow, only the cooling source moves through the material and domains form in spatial regions where the temperature is below the critical temperature. 
In the presence of flow, these domains get transported as well. 
The subsequent dynamics is important as it reflects how a moving fluid transports the phase-segregated structures that are formed by a moving cooling spot while phase separation is still going on. 
The latter non-equilibrium dynamics is rather complex because of the presence of time-dependent temperature field, time-dependent concentration field and fluid flow.  

At the continuum level, we employ phase field modeling to investigate the local phase separation under anisotropic conditions generated by a Gaussian cooling source moving through the two-component fluid. 
The phenomenological Cahn–Hilliard–Cook framework \cite{cahn1958,cook1970} is modified appropriately in order to take into account the spatially varying temperature field and its coupling to order parameter field. 
Further, to examine the influence of advective flow on phase separation coupling between flow velocity $\mathbf u$ and order parameter $\psi$ is incorporated via $\mathbf u \cdot \nabla \psi$.

The chosen system is scientifically interesting because of heterogeneous `aging' times. 
Unlike conventional isothermal bulk phase separation \cite{gerhard2024, subir-singdha2025, roynalina, roycoalescence, Tanaka1996, roy2013effects} or stationary thermal gradients \cite{lee2002,bertei2019}, here the ‘moving’ cooling source continuously generates regions with different quench histories along its direction of motion. 
Regions ahead of the cooling source remain unquenched, regions immediately behind the source have only recently entered the phase-separating regime, while regions farther behind have undergone coarsening for longer times. 
As a result, the system simultaneously contains domains of different ages. 
To the best of our knowledge, systems with inhomogeneous 'aging' times are not well studied. 

Without flow, the morphological evolution is governed by the competition between the cooling-source velocity and thermal diffusion. 
For very low source velocities, the system develops a nearly symmetric pattern. 
As the source velocity is increased, this symmetry is progressively broken, resulting in the formation of an asymmetric spiral-like structure with vertical layers formed to the right side of spiral as the time progresses, thus indicating a growing directional influence of the source motion on the domain morphology. 
Upon further increasing the source velocity, the spiral structure is no longer sustained, the thermal front lags behind the moving source, shortening the time available for domain growth and resulting in narrower, elongated structures. Thus, increasing the source velocity drives a systematic morphological transition from symmetric structures to asymmetric patterns. 

The moving cooling source continuously creates new phase-separated regions which have their own preferred directions of layer formation. 
The imposed flow velocity transports and modifies the already formed domains. 
As $u_x$ increases, this effect becomes stronger. 
The competition between phase separation and flow determines the final pattern morphology.
For low fluid velocity, horizontal layers develop in the downstream region, indicating the influence of material advection on the evolving phase-separated structure. 
And the region upstream retains distinct c-shaped structures and vertically oriented layers, reflecting the morphology established during the earlier stages of phase separation.
For an increase fluid velocity, layers predominantly align along the direction of fluid flow at all times. 
This is nicely captured in the anisotropy parameter $A$. 
The sign of $A$ characterizes the preferred orientation of the domains: $A<0$ corresponds to predominantly horizontal layers, whereas $A>0$ indicates predominantly vertical layers. 
For an isotropic morphology, the contributions from the two directions are comparable, giving $A=0$. 
Thus, the evolution of $A$ provides a quantitative measure of the transition from vertically oriented or isotropic structures to flow-aligned horizontal layers.
Note that in this paper, we consider the material flow by adding the $\mathbf u \cdot \nabla \psi$ term in \cref{chc2}, but we do not consider advection of the temperature field. 

The evolution of the domain morphology in the presence of material flow can be viewed in terms of two distinct spatial regions:
\begin{itemize}
    \item \textbf{Flow-aligned domain region:}
    The material has undergone phase separation for a longer period and got sufficient time to align the domain layers along the flow direction (horizontal).
     
    \item \textbf{Transverse domain region:}
    The domains/layers are predominantly vertical or C-shaped with respect to the flow direction.
    In this region, the morphology is still governed by the cold thermal front and influence of flow velocity is small.
   
\end{itemize}

Thus, the final microstructure is determined not only by the cooling temperature front but also by the competition between the thermal front and the material flow. 
A moving thermal history can encode a spatial record of non-equilibrium phase evolution, allowing controlled selection of material morphology through the competition between front propagation and transport. 
Our work is relevant for directional solidification, thermally driven manufacturing processes, printing, and coating where local phase evolution occurs under a spatially varying processing history. 

\vskip 1.0cm
Acknowledgement: SR thanks the Anusandhan National Research Foundation (ANRF) Advanced Research Grant (ARG) program (ANRF/ARG/2025/008570/PS) for the financial aid for this work. 
SR also acknowledges financial support from BITS Pilani, India, through NFSG grant no: N5/25/1053.

\bibliography{phasesep}
\end{document}